\documentclass[times,showpacs,onecolumn,superscriptaddress]{elsarticle}
\biboptions{numbers,sort&compress}
\usepackage{amssymb,mathrsfs,amsmath}
\usepackage{graphicx}
\usepackage{color}
\usepackage{ulem}
\usepackage{CJK}
\usepackage{lineno}

\journal{Journal of Sound and Vibration}

\begin{document}


\begin{frontmatter}

\title{Occurrence of gradual resonance in a finite-length granular chain driven by harmonic vibration}

\author[NJUST]{Tengfei Jiao} \author[NJUST]{Shutian Zhang} \author[NJUST]{Min Sun}
\author[NJUST]{Decai Huang \corref{Huang}}

\cortext[Corresponding author:] {hdc@njust.edu.cn}

\address{Department of Applied Physics, Nanjing University of Science and Technology, Nanjing 210094, China}

\date{\today}

\begin{abstract}

  This study presents numerical simulations of the resonance of a finite-length granular chain of dissipative grains driven by a harmonically vibrated tube. Multiple gradual resonant modes, namely, non-resonance mode, partial-resonance mode, and complete-resonance mode, are identified. With a fixed vibration frequency, increased vibration acceleration leads to a one-by-one increase in the number of grains participating in resonance, which is equal to the number of grain-wall collisions in a vibration period. Compared with the characteristic time of the grain-grain and the grain-wall collisions, the time of free flight plays a dominant role in grain motion. This condition results in the occurrence of large opening gaps between the grains and independent grain-grain and gain-wall collisions. A general master equation that describes the dependence of the system energy on the length of the granular chain and the number of grain-wall collisions is established, and it is in good agreement with the simulation results. We observe a gradual step-jump increase in system energy when the vibration acceleration is continuously increased, which is dedicated to an individual energy injection. Moreover, two typical phase diagrams are discussed in the spaces of $\phi-\it{\Gamma}$ and $N-\it\Gamma$.

\end{abstract}


\end{frontmatter}

\section{Introduction}

One-dimensional ordered granular grains exhibit rich nonlinear dynamical behaviors\cite{NestBook,Sen2008PR462.21}. The mechanical characteristic of granular grains can be tuned from having strong nonlinearity without static precompression to having weak nonlinearity with large static precompression. Many unique characteristics, including localized solitary wave, shock energy trapping, and wave filtering have been revealed\cite{Nest2006PRL96.058002, Daraio2012PRE85.037601,Yule2015NJP17.023015,Chong2017JSV393.216}. Similar phenomena in many physical settings, including nonlinear optics, and ordered crystal arrays, have also been attracted extensively research attention\cite{Chen2022JSV529.116966,Flach2008PRep467.1,Flach2019PRL122.054102, Zhang2012OE20.27888,Xu2016OL41.2656,Perego2021PRA103.013522,Sen2009APL95.224101,Sen2011APL99.063510}. Hence, further understanding the physical mechanisms of an aligned grains for controlling energy propagation, localization, and resonance could lead to new mechanical filtering and protective devices.

The nonlinear responses of granular chains mainly stem from two factors: nonlinear interaction between contacting grains because of the macroscopic size, and absence tensionless behavior for dispersed grains when they lose contact\cite{NestBook,Sen2008PR462.21,Sen1999PA274.188,Coste1997PRE56.6104,Sen2020chaos30.043101}. The describing equations of granular chains are non-integrable due to combined actions, and basic nonequilibrium and energy transport properties cannot be analytically solved. Pioneering studies were conducted by Nesterenko et al. on the propagation of a pulse in a granular chain of spherical grains aligned on a line\cite{Nest1983}. In the absence of static precompression, a harmonic acoustic wave cannot exist in a granular chain because an opening gap occurs between the adjacent grains and the acoustic velocity is zero, which is also called a sonic vacuum. When high precompression is applied, long wavelength approximation is used, and an analytical KdV solution is derived to model the propagation of the solitary wave. When the solitary wave is placed within two static walls, dispersion occurs, leading to multiple interactions between the dispersed solitary wave and the secondary waves\cite{Sen2011PRE84.046610,Sen2014PRE89.0532020,Sen2012EPL100.24003}.
At moderate precompression, the counterbalance between dispersion and nonlinearity results in the long-lived existence of the solitary wave.
Meanwhile, similar to the situation in a classical thermal system, a steady quasi-equilibrium state is eventually formed after sufficient time.
The initial impulse energy is equally shared by all grains of the granular chain, with a normal Gaussian distribution of grain velocities\cite{Sen2004PA342.336,Sen2015PRE91.042207,Sen2017PRE95.32903}.

In reality, granular materials are inherently finite in size and dissipative in collision. Energy has to be continuously injected by using an external excitation. Linear resonance is an intrinsic characteristic of an ordered linear system under harmonic excitation. Fundamental resonance occurs at the excitation frequency. The nonlinear responses of various systems to external excitation have been extensively investigated in past decades\cite{Gamma1998RMP70.223, Rajasekar2016Book,Vincent2020RSA379.20200226}. Examples include one-dimensional ordered systems, such as the Klein-Gorden chain\cite{Kovaleva2016PRE94.022208,Kovaleva2020PD402.132284}, the Fermi-Pasta-Ulam-Tsingou chain\cite{Pierangeli2018PRX8.041017,Kovaleva2018PRE98.052227,Berman2005Chaos15.015104}, and the Toda-lattice chain\cite{TodaBook,Vincent2018PRE98.062203,Vincent2021PD419.132853}. Recently, extensive attentions have been devoted to the dynamical responses of granular systems with external harmonic excitation\cite{Luding2013NJP15.113043,Poschel2015PRA3.024007, Sun2007PRE75.061302,Jia2019PRE99.042902}. For a homogeneous granular chain, the propagation band and attenuation bands depend on the amplitude and frequency of driven force. A notable characteristic is that the most effective energy transport occurs at the resonant frequencies\cite{Daraio2013PRE88.012206,Daraio2015PRE91.023208,Daraio2016PRE93.052203, Pozharskiy2015PRE92.063302,Zhang2017Exp57.505,Kovaleva2018PRE98.012205}. Unlike in an integrable system, the opening gap between grains is considered as one of the key factors for nonlinear responses. To our knowledge, only a few studies have been conducted on harmonically vibrated granular chains with large opening gaps\cite{Kadanlof1995PRL74.1268,Senf1995PRL74.2686,Luding1994PRE50.4113}. Further detailed studies on the homogeneous granular chain driven by harmonic vibration would be useful for an improved understanding of nonlinear responses.

The current study provides a simulation of a granular chain placed in a harmonically vibrating tube. Two synchronously vibrational boundary walls are regarded as external energy sources to keep the system in motion. Section II do this paper describes the simulation model and used parameters. Section III shows the detailed oscillatory motions of grains and multiple nonlinear resonant modes, namely, non-resonance mode(NRM), partial-resonance mode(PRM), and complete-resonance mode(CRM). Section IV presents the theoretical analyses of the characteristic times of grain-grain collision, grain-wall collision and free flight. A general master equation is proposed to describe the gradual resonant modes. Then, the dependence of system energy on vibration acceleration and two typical phase diagrams in the spaces of $\phi-{\it{\Gamma}}$ and ${N-\it{\Gamma}}$ are plotted. Section V summarizes the conclusions.

\section{Simulation model}

\begin{figure}[htbp]
\centering
  \includegraphics[width=0.5\textwidth, trim=4cm 13.2cm 3.5cm 12.5cm, clip]{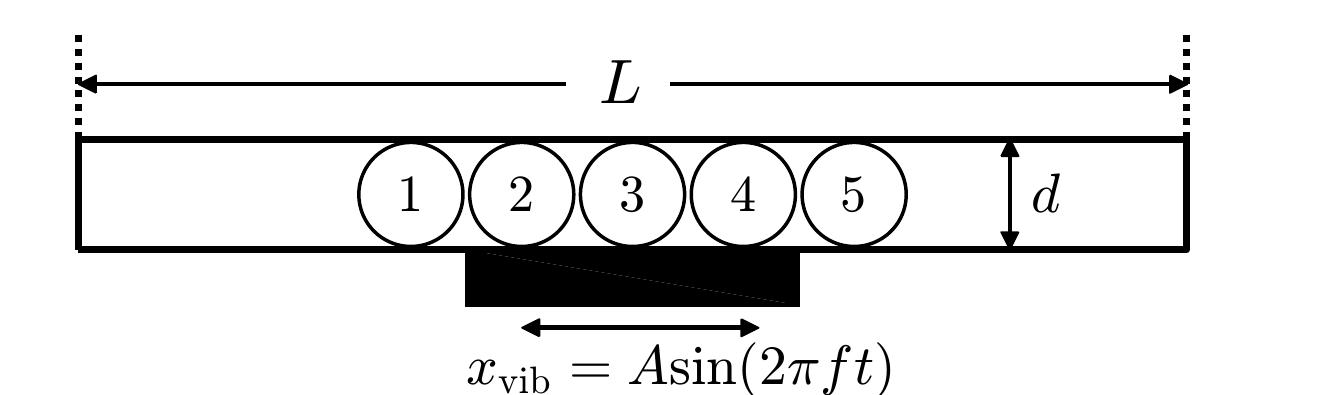}
  \caption {Schematic of $N$ identical spheres confined in a cylindrical tube. $L$ is the tube length. $A$ and $f$ denote the vibration amplitude and frequency, respectively.}
\label{fig:FigureModel}
\end{figure}

We consider horizontally aligned $N$ identical grains with diameter $d=5.0{\rm~mm}$ as shown in Fig.\ref{fig:FigureModel}. The grains are placed in a cylindrical tube, which exactly restrict the grains moving forth and back in the $x$-direction. The filling fraction is defined as $\phi=L_g/L$, where $L_g=Nd$ and $L$ is the tube length. The length of the clearance is $L_c=L-Nd$. The walls of the tube are infinitely smooth, that si, $\mu=0$. The tube is placed on a horizontal shaker with a displacement $x_{\rm{vib}} = A\sin(2\pi ft)$, indicating that the two boundary walls harmonically vibrate in phase. The corresponding dimensionless vibration acceleration is ${\it\Gamma} = A(2\pi f)^2/g$, with vibration amplitude $A$ and frequency $f$. The vibration period is $T=1/f$, and $g$ denotes gravitational acceleration.

The discrete element method is used in the numerical simulation. In each simulation time step $dt$, the positions and velocities of grains are updated by adopting Newton's second law of motion.
Only the translational motion is considered in the simulations. The model has no gravity. The force between two contacting grains is calculated only in normal directions with the Hertzian model as follows\cite{Lanndau1959,PRE2006Huang,PRE2012Huang,PT2022Huang,PLA2017Huang}.
\begin{equation}
F_{ij}=\begin{cases}
\eta k_{n}{\delta_{ij}}^{3/2} & \text{ , } \delta_{ij}> 0 \\
0 & \text{ , } \delta_{ij}< 0
\end{cases}
 \label{HertzFn}
\end{equation}

\noindent where $i$ and $j$ are the indexes of the adjacent grains. $\delta_{ij}=\max(0, d_{ij}-\left |{x}_i-{x}_j \right |)$ is the overlap of the contacting grains $i$ and $j$, and $d_{ij}$ is the sum of radiuses, $R_i$ and $R_j$, of the contacting grains. $k_n$ and $\eta$ characterizes the stiffness and damping of the grains,respectively, and are related to the materials of the contacting grains.
\begin{equation}
k_n=\frac {4}{3} \frac {{Y_i}{Y_j}}{Y_i+Y_j}\sqrt{\frac {{R_i}{R_j}}{R_i+R_j}}
\label{KN}
\end{equation}

\noindent where $Y=E/(1-{\nu}^2)$. $E$ is the Young's modulus, and $\nu$ is the Poisson ratio. The parameters of granular materials are listed in Table I. In the simulations, a certain energy dissipation is necessary to maintain the balance of the continuous energy injection. $\eta$, a dissipation coefficient, is introduced; its value is $1$ when two grains are in the compression process, and its value is $0.9$ when two grains are in the separation process. Thus, the restitution coefficient is actually $\eta=0.9$ for a collision event. The collisions between the outermost grains and the boundary walls are regarded as grain-grain collision, but that the walls have an infinite mass and diameter.
\begin{center}
{\bf{Table I}} Simulation parameters \\
\begin{tabular} { p{5.8cm} p{1.45cm} p{0.95cm} }
   \hline
   Quantity & Symbol   & Value \\
   \hline
   Grain diameter [mm] & \emph {$~~~R$} & 5 \\
   Density [$\rm {10^{3}kg/m}^{3}$] & \emph {$~~~\rho$} & 7.8 \\
   Young's modulus [GPa] & \emph {~~~E} & 10.0 \\
   Poisson ratio & $~~~\nu $ & 0.3 \\ 
   Simulation time-step [s] & $~~~dt$ & $10^{-8}$ \\
   \hline
\end{tabular}
\label{TableGrainPara}
\end{center}

At the beginning time $t=0$, all grains are placed on a line and barely touched one another. The grains are at rest, and the leftmost grain comes in contact with the left boundary wall only. External energy is injected by the collisions of the leftmost and rightmost grains with the corresponding boundary walls, and a certain fraction of energy is dissipated during the grain-grain and grain-wall collisions. In our simulation, the primary $500$ vibration periods is sufficient to achieve a steady vibration state, which means a dynamical equilibrium exists between the injection and dissipation energies. Then, the simulation is continued, and the calculations are performed in the subsequent simulation of $500$ periods.
\begin{figure}[htbp]
\centering
\includegraphics[width=0.45\textwidth, trim=3.7cm 8.3cm 5.8cm 7.5cm, clip]{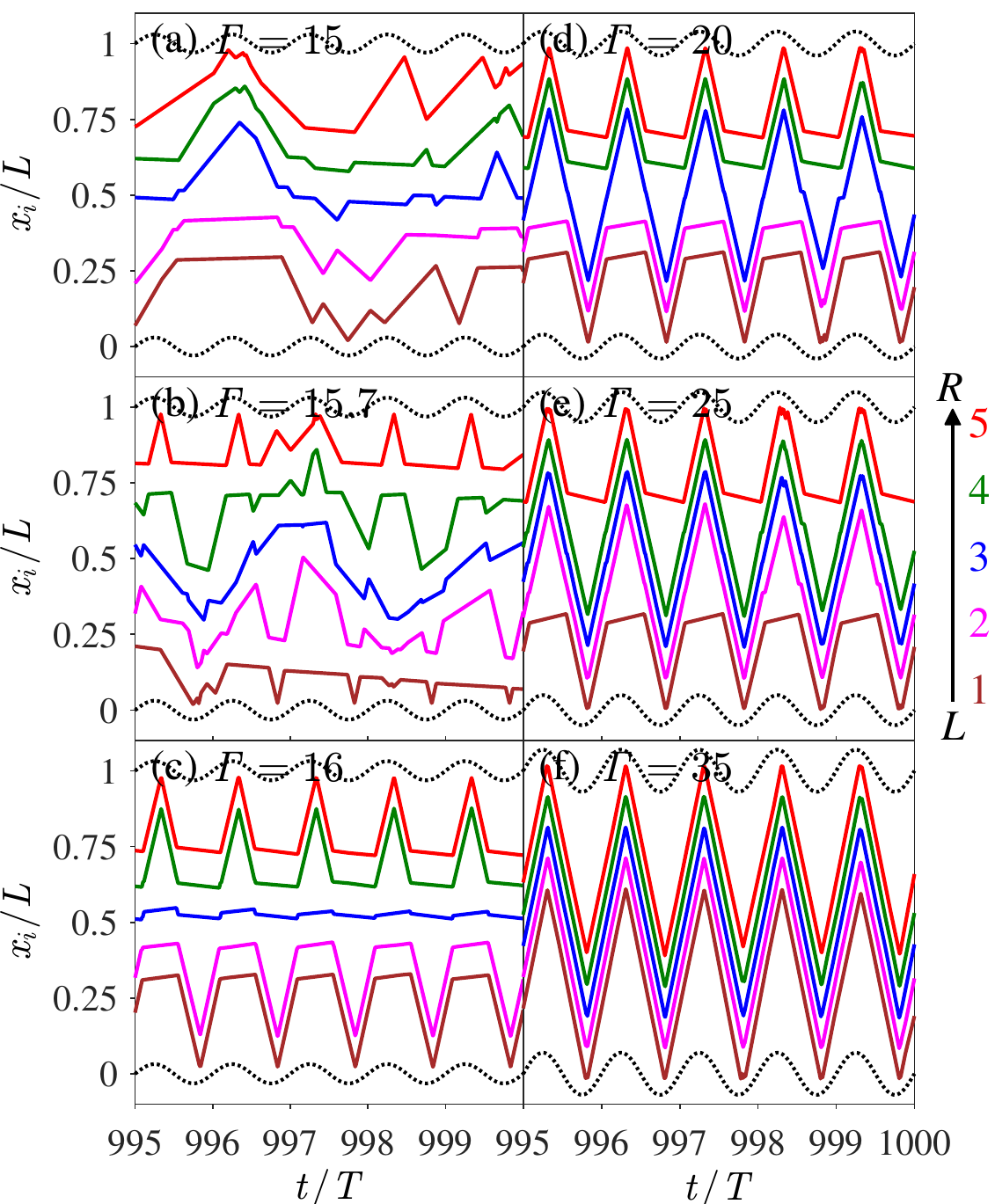}
  \caption {Temporal evolution of grain position at five vibration accelerations as indicated in the legend. The vibration frequency is fixed at $f=50{\rm~Hz}$. The solid lines from the bottom to the top in each panel represent the grains from the left $i=1$ to the right $i=5$, respectively. The bottom and top dashed lines denote the positions of the left and right boundary walls, respectively.}
\label{fig:FigPosition}
\end{figure}
\section{Gradual resonant modes} 
The simulation is performed first at a weak vibration acceleration, namely, ${\it\Gamma}=15$. The frequency is set to $f=50{\rm~Hz}$. The temporal evolutions of the center positions of the grains are plotted in Fig.\ref{fig:FigPosition}(a). Grain motion has no clearly regular dependence on the vibration mode of the tube. This disordered motion is thus named NRM. When the vibration acceleration is increased, i.e., ${\it\Gamma}=15.7$, the central grains ($i=2,3,4$) appear to move in NRM as shown in Fig.\ref{fig:FigPosition}(b). However, the outermost grains $(i=1,5)$ start to resonate, where five quasi-resonance period motions are clearly seen, and PRM occurs.
In each oscillatory period, the motion is composed of a peak-like part and a plain-like part.
The left and right resonant grains rapidly approach the corresponding walls in turn, but they fall behind the vibration motions of the corresponding boundary walls.
In addition, due to the disordered motion of the central grains, the amplitudes of the peaks are not completely equal, and a few disorder motions of the resonant grains occasionally happen, i.e., $i=1$ around time $t=996T$ and $i=5$ around time $t=997T$.

When the vibration acceleration continues to increase, that is, ${\it\Gamma}=16$, the system is still in the PRM and the number of resonant grains increases as shown in Fig.\ref{fig:FigPosition}(c). The leftmost and rightmost grains are forced to symmetrically resonate. They alternately move forth and back in each vibration period, which also have two resonant processes, a peak-like part and a plain-like part. However, compared with the resonant results in Fig.\ref{fig:FigPosition}(b), the occupied region of the peak in Fig.\ref{fig:FigPosition}(c) obviously extends. In addition, it is interesting that an almost flat line appears in the middle of Fig.\ref{fig:FigPosition}(c), indicating that grain $3$ {\em stays} around the center region of the tube throughout the vibration period.

As shown in Fig.\ref{fig:FigPosition}(d), the vibration acceleration is set to ${\it\Gamma}=20$. Similar resonant peaks and plains as those in Fig.\ref{fig:FigPosition}(c) are observed. However, all grains are involved in the resonance in each vibration period. The peak-like resonant process involves three grains. The plain-like resonant pattern completely disappears in the central grain $i=3$.
This result means that grain $i=3$ starts to enter the complete resonant pattern, and the total number of grains in the peak-like resonance is $n=3$.
Further increasing the vibration acceleration increases the number of grains participating in complete resonance, as shown in Fig.\ref{fig:FigPosition}(e), i.e., ${\it\Gamma}=25$.
The peak-like resonant pattern spreads from the central grain to the ends one by one.
When $\it{\Gamma}$ is increased adequately, i.e., ${\it\Gamma}=35$, all grains resonate completely in the peak-like pattern($n=5$), and CRM occurs as shown in Fig.\ref{fig:FigPosition}(f).
They oscillate like a whole ensemble synchronously resonating with the tube.
The granular chain collides with one of the boundary walls and rebounds back towards the other one through a transient collision.
After a time of free flight across the tube, the granular chain arrives at the opposite wall, and a similar process occurs.
This round trip is of in the same period as that of the tube vibration.

\begin{figure}[htbp]
\centering
\includegraphics[width=0.42\textwidth, trim=4.4cm 9.5cm 5cm 10.5cm, clip]{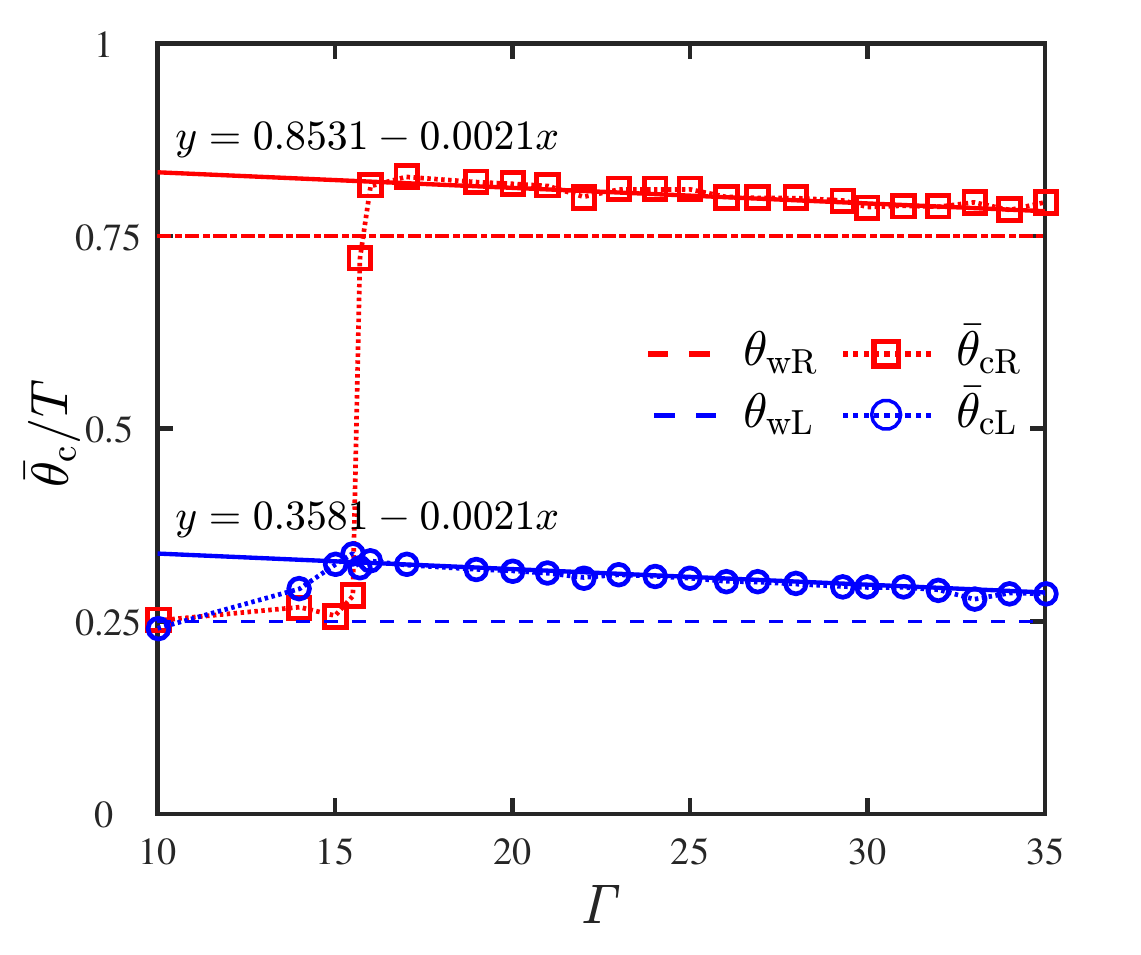}
  \caption{(Color online.) Collision phase as a function of vibration acceleration at a fixed vibration frequency $f=50~\rm{Hz}$. The red dashed-dotted and blue dashed lines indicate the phase when the left and right walls reach the farthest position, respectively. The red squares and circles denote the phase lag between the outmost grains and the right and left walls, respectively. The solid lines are obtained by using the fitted equations $y=0.3581-0.0021x$ and $y=0.8531-0.0021x$, respectively.}
\label{fig:FigPhaseDiff}
\end{figure}

In Fig.\ref{fig:FigPosition}, an obvious phase lag occurs, which means that the outermost grain has not reached the position of the peak when the corresponding wall reaches its farthest position. Here, the collision phase $\theta_{\rm{c}}$ is defined as the moment when the collision force between the outermost grain and the wall reaches the maximum in a vibration period.
For the vibrating tube, when the right and left walls reach their farthest positions, the corresponding phases are $\theta_{\rm{wR}}=\pi/2$ and $\theta_{\rm{wR}}=3\pi/2$, respectively. In Fig.\ref{fig:FigPhaseDiff}, the time-averaged collision phase as a function of vibration acceleration is plotted at a fixed vibration frequency of $f=50~\rm{Hz}$. When $\it{\Gamma}$ is larger than $16$, the collision phases of $\bar{\theta}_{\rm{cL}}$ and $\bar{\theta}_{\rm{cR}}$ are always larger than the corresponding phases of the left and right walls respectively, which is consistent with the results in Fig.\ref{fig:FigPosition}.
Meanwhile, an increase in vibration acceleration leads to a linear decrease in the collision phase. The two fitting equations have the same linear fitting coefficient due to the symmetry of the vibration.

\begin{figure}[htbp]
\centering
\includegraphics[width=0.47\textwidth, trim=3.2cm 9.2cm 3.3cm 7.7cm, clip]{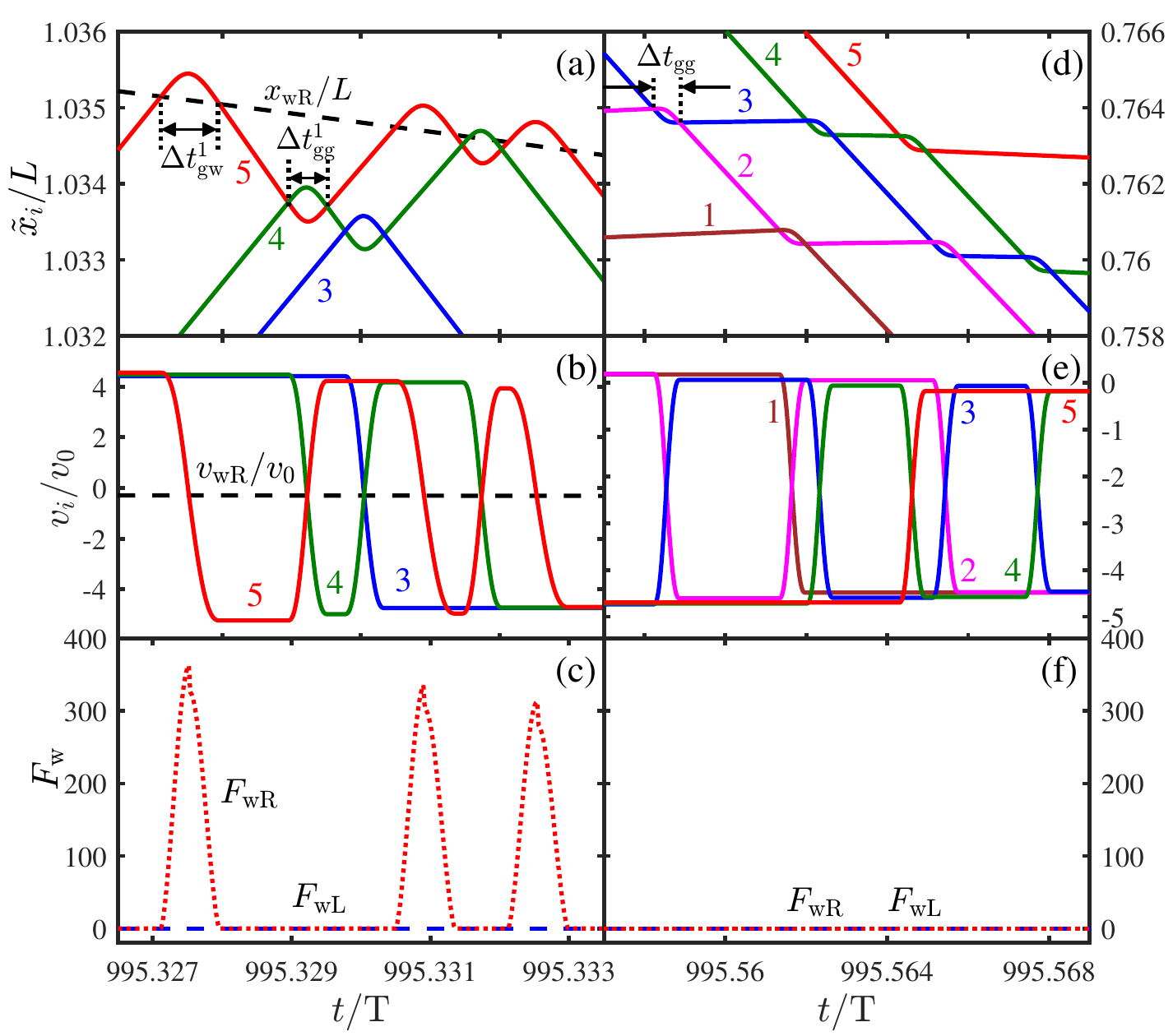}
\caption{(Color online.) Temporal evolution of (a)(d)grain position, (b)(e)grain velocity, and (c)(f)collision force applied on the boundary walls. The results are extracted from Fig.\ref{fig:FigPosition}(d). In (a), the positions are shifted with $\tilde{x}_5=x_5+R$, $\tilde{x}_4=x_4+3R$, and $\tilde{x}_3=x_3+5R$. In (a)(b), the dashed lines are the position and vibration velocity of the right wall. The grain index, the collision time, and the collision force are marked in each panel.}
\label{fig:FigXVF}
\end{figure}

When the external supplied energy obtains a balance with the energy loss, the system achieves a steady resonant state.
A vibration period has two typical collision cases, namely, chain-wall collision and chain-chain collisions.
The former occurs in the regions of two boundary walls, and the latter happens around the central region of the tube.
With the condition of ${\it{\Gamma}}=20$ for an example, Figs.\ref{fig:FigXVF}(a)(b)(c) show the temporal evolution of the dynamical behaviors of the granular chain for the chain-wall collision, and Figs.\ref{fig:FigXVF}(d)(e)(f) pertain to the chain-chain collision.
The results in Fig.\ref{fig:FigXVF}(a) are the enlarged part in the beginning of period $995$ in Fig.\ref{fig:FigPosition}(d). In Fig.\ref{fig:FigXVF}(a), grains $5,4$ and $3$ approach the right wall, and large opening gaps are observed between the adjacent grains.
The straight lines suggest that the grains fly freely before the collision happens.
Grain $5$ as the rightmost one first collides with the right boundary wall. The intersection regions between the black dashed line and the red solid line indicate the occurrence of grain-wall collision. The first labelled collision time is $\Delta{t}_{\rm{gw}}^{1}=0.00082T$. Then, it bounces back and undergoes a short time of free flight. The first grain-grain collision occurs between grain $5$ and grain $4$, whose time of duration is $\Delta{t}_{\rm{gg}}^{1}=0.00056T$. They turn around after the collision and run into the next free-flight motion. Grain $5$ collides with the right wall again, and grain $4$ collides with grain $3$. Grain $4$ and grain $3$ exchange their motion directions; the former continues to fly freely to the right wall, and the latter flies to the left wall. Grain $4$ collides with the rebounded grain $5$, and they exchange their motion directions again. When grain $5$ completes its third collision with the right wall, the chain-wall collision ends.
Afterwards, grains $5,4$ and $3$ fly freely toward the left wall and collide with grains $2$ and $1$.
During the chain-wall collision, the total number of collisions between grains is ${\sum_{1}^{n-1}} n_{i}$.

In Fig.\ref{fig:FigXVF}(b), the temporal reduced grain velocity $v_i/v_0$, where $v_0=2{\pi}Af$, is plotted for grains $3,4$, and $5$, in which the same oscillatory characteristic is reproduced. Grain $5$ experiences three grain-wall collisions with the right wall. Each collision happens at the moment of $v_{\rm{wR}}/v_0\neq0$, which confirms that the phase lag is consistent with the results shown in Fig.\ref{fig:FigPhaseDiff}. Two grain-grain collisions occurs for grain $5$ and grain $4$, but only one grain-grain collision happens between grains $4$ and $3$. In each grain-wall collision, the rebounded velocity after collision is larger than the incident velocity before collision. The external energy is injected into the system. Meanwhile, in each grain-grain collision, the exchanged velocities after collision are lower than the incident velocities before collision. This finding implies that a certain amount of energy is dissipated.

Fig.\ref{fig:FigXVF}(c) shows the temporal evolution of collision force between the outermost grain with the boundary wall.
Three independent peaks are seen for the right wall labelled by $F_{\rm{wR}}$, which is equal to the number of grains in the peak-like pattern $n$, and no collision happens for the left wall labeled by $F_{\rm{wL}}$ during the right chain-wall collision. The time of the grain-wall collision of $F_{\rm{wR}}$ has the same order as that of grain-grain collision.

In Fig.\ref{fig:FigXVF}(d), the results are also the enlarged parts of Fig.\ref{fig:FigPosition}(d) after the first peak of period $995$. All grains are located around the position of $3L/4$. Grains $2$ and $1$ as the right-traveling ones collide with grains $3,4$ and $5$ rebounded from the right wall.
At this chain-chain collision stage, the adjacent grains experience a certain number of grain-grain collisions $n(n-1)$.
Compared with $T$, a single grain-gain collision is rather small, i.e., $\Delta{t}_{\rm{gg}}=0.00068T$ as shown in Fig.\ref{fig:FigXVF}(d). Similar to the right chain-wall collision in Fig.\ref{fig:FigXVF}(a), all grains move to the left wall after the last collision between grain $4$ and grain $5$. In Fig.\ref{fig:FigXVF}(e), the same properties of grain-grain collisions are repeated for grain velocity.
The grain velocity after collision is lower than that before collision because of the dissipation. Meanwhile, every grain has a negative value of the velocity, which represents the left-traveling motion. Grains $1,2$ and $3$  have obviously higher velocities than grains $4$ and $5$, whose velocities are low. As indicated in Fig.\ref{fig:FigXVF}(f), no grain-wall collisions occur during these chain-chain collisions happen.

When the collision between grains $4$ and $3$ is completed, all grains start to fly freely to the left wall. Similar to the chain-wall collision shown in Fig.\ref{fig:FigXVF}(a)(b)(c), grain $1$ as the leftmost incident grain first collides with the left wall.
A series of collisions occur, comprising $n=3$ times of grain-wall collisions between grain $1$ and the left wall, and ${\sum_{1}^{n-1}} n_{i}$ times of grain-grain collision between the grains of $1,2,3$.
Afterward, they rebound back and collide $n(n-1)$ times together with grains $4$ and $5$, similar to the chain-chain collision shown in Fig.\ref{fig:FigXVF}(d)(e)(f).
Grains $3,4$ and $5$ acquire high velocities and fly to the right wall. When the collision between grain $5$ and the right wall occurs, a whole oscillatory motion of the granular chain is completed. Then, the next periodical oscillatory motion starts.

\section{Discussions}

 On the basis of the above-mentioned observations, instead of forming a homogeneous quasi-equilibrium state\cite{Sen2004PA342.336,Sen2015PRE91.042207,Sen2017PRE95.32903}, three resonant modes, namely, NRM, PRM, and CRM, are identified in turn with the increase in ${\it{\Gamma}}$. In PRM, and CRM modes, two characteristic times, i.e., collision and free flight, determine the resonant dynamics of the granular chain. The former includes two kinds of collisions, namely, grain-grain collision and grain-wall collision. The ideal collision between a spherical grain and an identical one or a static wall is worthy of investigation. This situation means that both the grain and the wall are completely elastic. In this ideal condition, the grain-wall collision time is strictly derived based on Hertz's theory \cite{Lanndau1959,PLA2017Huang}.
\begin{equation}
\tau_{\rm{gw}}=2.94(\frac{4 k_{\rm{n}}}{5m})^{-2/5}v_{i}^{-1/5},
\label{eq:Timetaugw}
\end{equation}

\noindent where $v_{i}$ is the incident velocity of the grain and $m$ is the grain mass. For the grain-grain collision, the relative velocity of the grains is $v_{ij}$. The characteristic collision time is written as follows:
\begin{equation}
\tau_{\rm{gg}}=2.23(\frac{4 k_{\rm{n}}}{5m})^{-2/5}v_{ij}^{-1/5}.
\label{eq:Timetaugg}
\end{equation}

According to Eqs.(\ref{eq:Timetaugw})(\ref{eq:Timetaugg}), the theoretical results of $\tau_{\rm{gw}}$, and $\tau_{\rm{gg}}$ are in the same order as the simulation results of $\Delta{t}_{\rm{gg}}^{1}$ and $\Delta{t}_{\rm{gw}}^{1}$ presented in Fig.\ref{fig:FigXVF}. However, they are three orders of magnitude smaller than vibration period $T$ which  suggests that all grains mainly move back and forth in a free-flight state. Therefore, compared with the vibration period $T$, grain-grain and grain-wall collisions can be regarded as approximately instantaneous ones.
For a fixed $f$, the vibrational velocity of boundary wall $v_{\rm{vib}}=v_0\cos(\omega t)$ is directly proportional to $\it{\Gamma}$. The velocity of the rebounded grain should also increase linearly with an increase in $\it{\Gamma}$.
Therefore, the resonated grain can be approximately regarded as free flying from one side of the tube to the other side because of the transient grain-grain collision.
This condition leads to the appearance of a large opening gap during the long time of free flight.
Our previous studies have also shown that a train of oscillatory collisions between the outermost grain and the wall occurs when a granular chain collides with a wall, and these are rediscovered in Fig.\ref{fig:FigXVF}\cite{PLA2017Huang}.
In each grain-wall collision, the outermost grain gains a certain amount of extra energy. Meanwhile, a smaller amount of energy is lost due to the dissipative grain-grain collision. When the injected and dissipated energies achieve a dynamical balance, the system enters a stable resonant state.
\begin{figure}[htbp]
\centering
\includegraphics[width=0.47\textwidth, trim=3.5cm 9.8cm 3.9cm 9.5cm, clip]{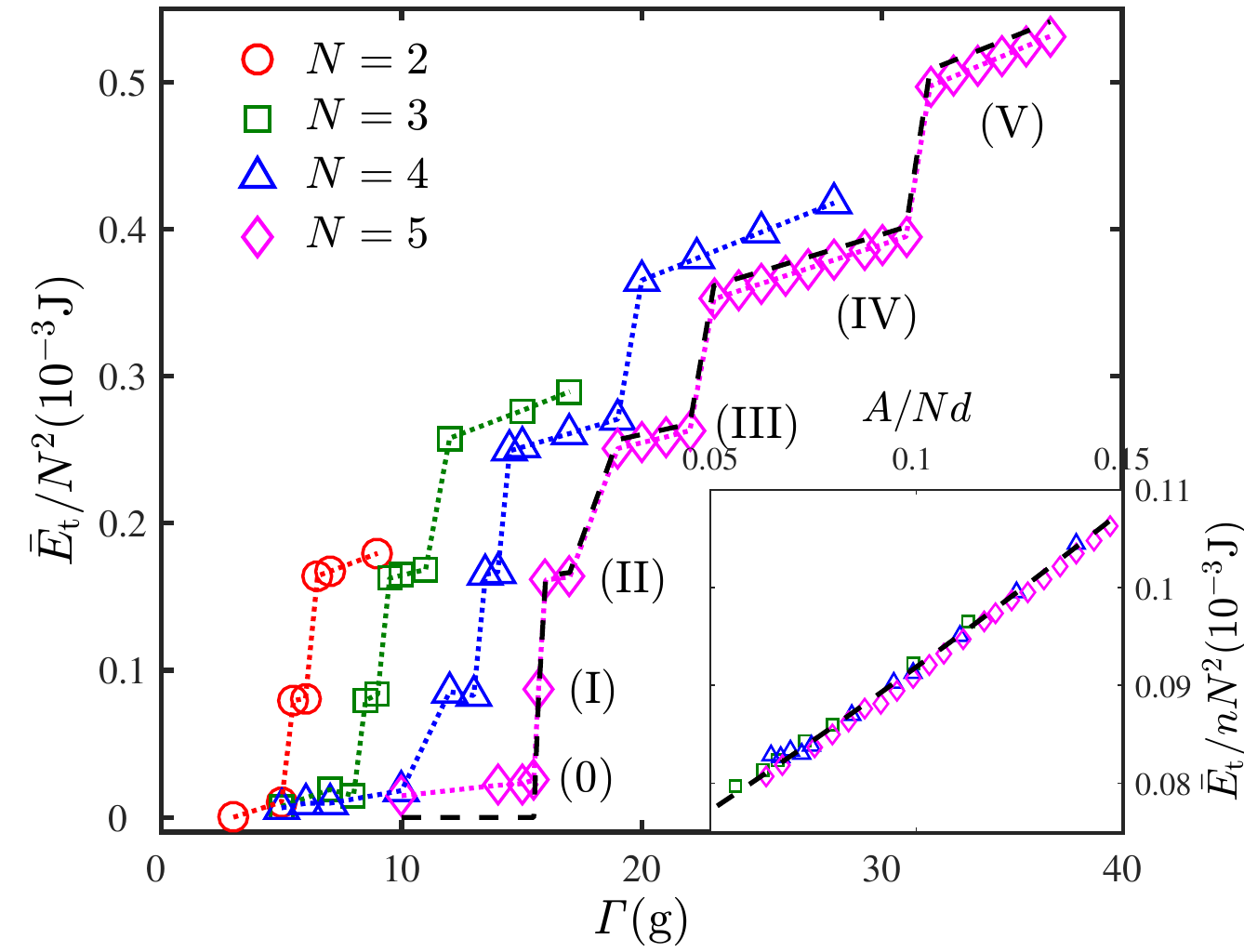}
  \caption{Reduced time-averaged energy of the granular chain as a function of the vibration acceleration $\it{\Gamma}$. The circles, squares, upwards triangles and diamonds denote chain lengths having $N=2,3,4$, and $5$ grains, respectively. The dotted lines are visual guidance. The dashed lines show obtained results by using Eq.(\ref{eq:Et0}). $\rm{0,I,II,III,IV}$ and $\rm{V}$ represent the number of resonance grains. The inset shows the simulation results for the chain having resonance grains $N{\ge}1$.
  }
\label{fig:FigureEnJump}
\end{figure}

Clearly, the external energy is individually injected by the separated grain-wall collision between the outermost grain and the boundary wall.
The number of grain-wall collisions determines the final total energy of granular chain due to the occurrence of oscillatory grain-wall collision.
Furthermore, it decides the number of grains participating in resonance. For the vibrating tube, the dynamical clearance of the system is written as:
\begin{equation}
\begin{aligned}
L_{\rm{C}}&=L_{\rm{c}}+2A   \\
&=(1/\phi-1)NR+2A,
\end{aligned}
\label{eq:ClearanceL}
\end{equation}
Thus, the averaged free-flight velocity of a grain is approximately calculated as following
\begin{equation}
\bar{v}_{\rm{g}}=2L_{\rm C}f.
\label{eq:Velocity}
\end{equation}

In Eq.(\ref{eq:Velocity}), the time-averaged grain velocity $\bar{v}_{\rm{g}}$ is proportional to the vibration amplitude $A$. At a fixed vibration frequency $f$, $\bar{v}_{\rm{g}}$ is proportional to $\it{\Gamma}$. This characteristic is consistent with the linear relation between $\bar{\theta}_{\rm{c}}$ and $\it{\Gamma}$ shown in Fig.\ref{fig:FigPhaseDiff}. Next, we can calculate the time-averaged energy of granular chain $\bar{E}_{\rm{t}}$ by summing the kinetic energies of all resonated grains because the potential energy is mostly zero due to the opening gaps between the grains.
\begin{equation}
\begin{aligned}
\bar{E}_{t}&\approx n\bar{e}_k=nm\bar{v}^2_{\rm{g}}/2 \\
&=2nN^2mR^2f^2[(1/\phi-1)+2A/NR]^2. \\  
\end{aligned}
\label{eq:Et0}
\end{equation}

In Fig.\ref{fig:FigureEnJump}, the reduced time-averaged energy of granular chain $\bar{E}_{\rm{t}}/N^2$ as a function of vibration acceleration $\it{\Gamma}$ is plotted for $N=2,3,4$ and $5$.
The time interval between $900T$ and $1000T$ is accounted in the calculations. The filling fraction is fixed at $\phi=0.5$.
With the five-grain chain as an example, at the range of weak vibration, an increase in $\it{\Gamma}$ leads to a linear increase in $\bar{E}_{\rm{t}}$, which is labeled by mode $0$.
As $\it{\Gamma}$ increases, a gradual step-jump increase in $E_{\rm{t}}$ is observed and labeled by modes $\rm{I,II,III,IV}$ and $\rm{V}$.
Fig.\ref{fig:FigureEnJump} also shows that $\bar{E}_{\rm{t}}/N^2$ increases monotonously with an increase in $\it{\Gamma}$ between the two adjacent energy jumps.
Eq.(\ref{eq:Et0}) is used to fit the simulation results, and an expected agreement is obtained, as shown in the inset of Fig.\ref{fig:FigureEnJump}, which plots $\bar{E}_{\rm{t}}/nN^2$ versus $A/Nd$ for all granular chains.
Consistent with the theoretical analysis results, the simulation results collapse into a line, which proves the validity of our analysis.
A series of independent energies are injected into the granular chain by the separated grain-wall collisions.
Under the condition of fixed vibration frequency, an increase in vibration amplitude implies an increase in vibration velocity, which further leads to an increase in rebounded grain velocity because the boundary wall has an infinite mass.
In general, the occurrence of the gradual resonance phenomenon is an intrinsic characteristic of a harmonically vibrating granular chain.
With an increase in vibration acceleration, the resonant mode starts from NRM to PRM and finally reaches CRM, during which the grains are partially resonated from the sides one by one.
The number of gradual resonant modes is equal to the length of the granular chain.
As the gradual resonance occurs, the total energy of the granular chain exihits a step-jump increase.
\begin{figure}[htbp]
\centering
  \includegraphics[width=0.47\textwidth, trim=3.5cm 9.8cm 3.9cm 9cm, clip]{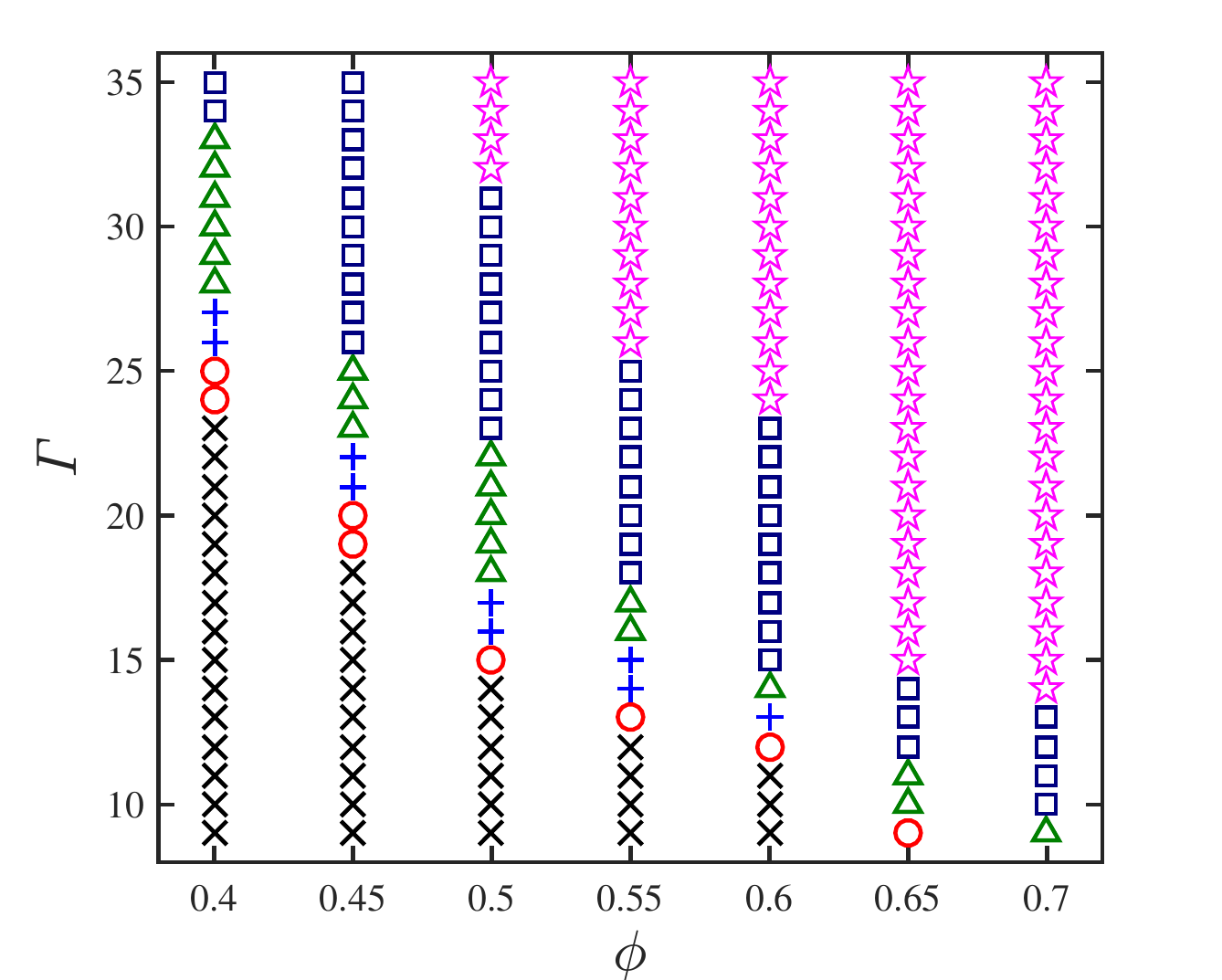}
  \caption {Phase diagram in the filling fraction $\phi$ and vibration acceleration $\it{\Gamma}$ space for the five-grain system. The symbols $\times,\bigcirc,+,\triangle,\square,\heartsuit$ represent the number of resonated grains for $0,1,2,3,4,5$, respectively.}
\label{fig:FigPhase}
\end{figure}

To provide a comprehensive understanding of the resonance of the granular chain, we plot a phase diagram in the filling fraction $\phi$ and the vibration acceleration $\it{\Gamma}$ space for the five-grain system, as shown in Fig.\ref{fig:FigPhase}. NRM appears in the left-bottom region and has low $\phi$ and $\it{\Gamma}$; meanwhile, CRM occurs in the right-top region and has high $\phi$ and $\it{\Gamma}$. PRM emerges in the crossover regions with moderate values of $\phi$ and $\it{\Gamma}$. Fig.\ref{fig:FigPhase} shows that the higher $\phi$ is and the larger $\it{\Gamma}$ is, the more the grains resonate.
\begin{figure}[htbp]
\center
  \includegraphics[width=0.47\textwidth, trim=3.5cm 9.0cm 3.9cm 8.5cm, clip]{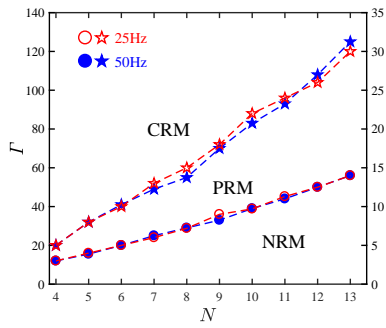}
  \caption {Phase diagram in chain length $N$ and vibration acceleration $\it{\Gamma}$ space for filling fraction $\phi=0.5$. The open and solid circles, and the stars indicate the boundaries of NRM, PRM and CRM, respectively, for vibration frequencies of $f=25$ and $50~{\rm{Hz}}$, respectively. The dashed lines are guides for the eye.}
\label{fig:FigPhaseGN}
\end{figure}

In Fig.\ref{fig:FigPhaseGN}, we plot the phase diagram of gradual resonant modes in chain length $N$ and vibration acceleration $\it\Gamma$ space. The filling fraction is set to $\phi=0.5$. NRM appears in the right-bottom region, and CRM emerges in the left-top region. These observations indicate that high $\it\Gamma$ and small $N$ lead to the occurrence of CRM, and low $\it\Gamma$ and large $N$ are required for the emergence of NRM. Again, PRM appears in the regions of moderate $\it\Gamma$ and $N$. When the vibration frequency increases, a small $N$ is favorable for the occurrence of CRM.

\section{Conclusions}

We numerically investigated the nonlinear resonance of a finite-length granular chain by using the discrete element method. The granular chain is placed in a harmonically vibrated tube and the external energy is injected into the system by the grain-wall collisions between the outermost grains and the boundary walls.

At a fixed vibration frequency, the simulational results show that the NRM occurs at small vibrational acceleration, in which the grain motion is in a disordered state and has no clear regular relation with tube vibration. When the vibrational acceleration is continuously increased, the number of grains participating in resonance increases one by one and PRM appears.
For a sufficiently large vibrational acceleration, all grains join in the resonance and CRM occurs. Compared with the characteristic times of grain-grain and grain-wall collisions, the free-flight time plays the dominant role in the oscillatory motion of the grains.
This feature results in a large opening gap between the grains and a phase lag between the resonance of the granular chain and the vibration of the tube. Further calculations reveal  the occurrence of independent oscillatory grain-grain and gain-wall collisions, and the number of gain-wall collisions in a vibration period is equal to the number of grains participating in resonance.

On the basis of the characteristics of the collisions and free flight, we posit that the external energy is individually injected into the system  by the independent oscillatory grain-wall collisions between the outermost grains and the boundary walls. A steady oscillatory state is achieved when the injected and dissipated energies reach a dynamical balance.
The linear dependence of the averaged free-flight velocity of grains on the product of vibration amplitude and vibration frequency is observed.
Moreover, we establish a general master equation of the total energy, including the length of granular chain and the number of grain-wall collisions.
The theoretical predictions are in good agreement with the simulation results.
As the vibrational acceleration increases, a gradual step-jump increase in energy is observed with the occurrence of the transition from NRM to CRM. Then, we find that a large filling fraction and high vibration acceleration are favorable for the occurrence of CRM. When the vibration frequency increases, decreasing the length of the granular chain is favorable for the occurrence of CRM.
The findings on gradual step-jump resonance in this study might be useful for designing new instruments of granular mechanical dampers.

\section*{Acknowledgments}

This work was supported financially by the National Natural Science Foundation of China (Grant No. 11574153).
\\
\noindent \textbf{DATA AVAILABILITY}

Data sharing is not applicable to this article as no new data were created or analyzed in this study.

\clearpage


\noindent

\begin{thebibliography}{9}
\bibitem{NestBook} V. Nesterenko, {\em Dynamics of heterogeneous materials} (Springer, New York, 2001).

\bibitem{Sen2008PR462.21} S. Sen, J. B. Hong, J. Bang, E. Avalos, and R. Doney, Phys. Rep. {\bf 462}, 21(2008).

\bibitem{Nest2006PRL96.058002} C. Daraio, V. F. Nesterenko, E. B. Herbold, and S. Jin, {\bf 96}, 058002(2006).

\bibitem{Daraio2012PRE85.037601} Y. Man, N. Boechler, G. Theocharis, P. G. Kevrekidis, and C. Daraio, Phys. Rev. E {bf 85}, 037601(2012).

\bibitem{Yule2015NJP17.023015} C. R. KWindows-Yule, A. D. Rosato, A. R. Thornton and D. J. Parker, New J. Phys. {\bf 17}, 023015(2015).
\bibitem{Chong2017JSV393.216} S. H. Chong, ang J. Y. Kim, J. Sound and Vib. {\bf 393}, 216(2017).
\bibitem{Chen2022JSV529.116966} J. G. Gui, T. Z. Yang, M. Q. Niu, and L. Q. Chen, J. Sound Vib. {\bf 529}, 16966(2022).

\bibitem{Flach2008PRep467.1} S. Flach, and A. V. Gorbach, Phys. Rep. {\bf 467}, 1(2008).

\bibitem{Flach2019PRL122.054102} T. Milthun, C. Danieli, Y. Kati, and S. Flach, Phys. Rev. Lett. {\bf 122}, 054102(2019).

\bibitem{Zhang2012OE20.27888} C. Zhao, Y. Zou, Z. T. Wang, S. B. Lu, H. Zhang, S. C. Wen, and D. Y. Tang, Opt. Express, {\bf 20}, 27888(2012).

\bibitem{Xu2016OL41.2656} G. Xu, A. Mussot, A. Kudlinski, S. Trillo, F. Copie, and M. Conforti, Opt. Lett. {\bf 41}, 2656(2016).
\bibitem{Perego2021PRA103.013522} A. M. Perego, A. Mussot,  and M. Conforti, Phys. Rev. A, {\bf 103}, 013522(2021).

\bibitem{Sen2009APL95.224101} R. P. Simon, A. Sokolow, and S. Sen, Appl. Phys. Lett.,{\bf 95}, 224101(2009).

\bibitem{Sen2011APL99.063510} A. Brendel, D. K. Sun, and S. Sen, Appl. Phys. Lett.,{\bf 99}, 063501(2011).

\bibitem{Sen1999PA274.188} M. Manciu,S. Sen, A. J. Hurd, Physica A, {\bf 274}, 588(1999). M. Manciu,S. Sen, A. J. Hurd, Physica A, {\bf 274}, 607(1999).

\bibitem{Coste1997PRE56.6104}C. Coste, E. Falcon, and S. Fauve, Phys. Rev. E {\bf 56}, 6104 (1997).

\bibitem{Sen2020chaos30.043101}G. Deng, G. Biondini, and S. Sen, Chaos {\bf 30}, 043101(2020).

\bibitem{Nest1983} V. Nesterenko, J. Appl. Mech. Tech. Phys. {\bf 24}, 733(1983); A.N. Lazaridi and V.F. Nesterenko, J. Appl.Mech. Tech. Phys. {\bf 26}, 405(1985); V.F. Nesterenko, J. Physique IV, {\bf C8}(1998).

\bibitem{Sen2011PRE84.046610} E. {\'{A}}valos, D. K. Sun, R. L. Doney, S. Sen, Phys. Rev. E {bf 85}, 046610(2011).
\bibitem{Sen2014PRE89.0532020} E. {\'{A}}valos, and S. Sen, Phys. Rev. E {bf 89}, 053202(2014).

\bibitem{Sen2012EPL100.24003} Y.Takato, and S. Sen, EPL {\bf 100}, 24003(2012).

\bibitem{Sen2004PA342.336} S. Sen, T. R. Krishna, and J. M. M. Pfannes, Physica A {bf 342}, 336(2004).

\bibitem{Sen2015PRE91.042207} M. Przedborski, T. A. Harroun, and S. Sen, Phys. Rev. E {bf 91}, 042207(2015).

\bibitem{Sen2017PRE95.32903} M. Przedborski, S. Sen, and T. A. Harroun, Phys. Rev. E {bf 95}, 032903(2017).
\bibitem{Gamma1998RMP70.223} L. Gammaitoni, P. H{\''{a}}nggi, P. Jung, and F. Marchessoni, Rev. Mod. Phys. {\bf 70}, 223(1998).

\bibitem{Rajasekar2016Book} S. Rajasekar, and Sanju{\'{a}}, {\em Nonlinear resonance} (Springer, Switzerland, 2016).

\bibitem{Vincent2020RSA379.20200226} U. E. Vincent, P. V. E. Mcclintock, I. A. Khovanov, and S. Rajasekar, Phil. Trans. R. Soc. A {\bf 379}, 20220226(2020).

\bibitem{Kovaleva2016PRE94.022208} A. Kovaleva, Phys. Rev. E {\bf 94}, 022208(2016).

\bibitem{Kovaleva2020PD402.132284} A. Kovaleva, Physica D {\bf 402}, 132284(2020).

\bibitem{Pierangeli2018PRX8.041017} D. Pierangeli, M. Flammini, L. Marcucci, G. Marcucci, and A. J. Agranat, Phys. Rev. X {\bf 94}, 041017(2018).

\bibitem{Kovaleva2018PRE98.052227} A. Kovaleva, Phys. Rev. E {\bf 98}, 052227(2018).

\bibitem{Berman2005Chaos15.015104} C. P. Berman, and F. M. Izrailev, Chaos {\bf 15}, 015104(2020).

\bibitem{TodaBook} M. Toda, {\em Theory of nonlinear lattices} (Springer, New York, 1988).

\bibitem{Vincent2018PRE98.062203} U. E. Vincent, T. O. Roy-Layinde, O. O. Popoola, P. O. Adesina, and P. V. E. Mcclintock, Phys. Rev. E {\bf 98}, 0362203(2018).
\bibitem{Vincent2021PD419.132853} O. Kolebaje, O. O. Popoola, and U. E. Vincent, Physica D {\bf 419}, 132853(2021).
\bibitem{Luding2013NJP15.113043} N. Rivas, S. Luding, and A. R. Thorton, New J. Phys. {\bf 15}, 113043(2013).

\bibitem{Poschel2015PRA3.024007} J. E. Kollmer, M. Tupy, M. Heckel, A. Sack, and T. P{\''{o}}schel, Phys. Rev. Appl. {\bf 3}, 024007(2015).

\bibitem{Sun2007PRE75.061302} Q. Shi, G. Sun, M. Hou, and Q. Lu, Phys. Rev. E {\bf 75}, 061302(2007).

\bibitem{Jia2019PRE99.042902} J. Brum, J. L. Gennisson, M. F., A. Tourin, and X. P. Jia, Phys. Rev. E {\bf 99}, 042902(2019).
\bibitem{Daraio2013PRE88.012206} J. Lydon, K. R. Jayaprakash, Y. L. Starosvetsky, A. F. Vakakis, and C. Daraio, Phys. Rev. E {\bf 88}, 012206(2013).

\bibitem{Daraio2015PRE91.023208}C. Chong, E. Kim, E. G. Charalampidis, H. Kim, F. Li, P. G. Kevrekidis, J. Lydon, C. Daraio, and J. Yang, Phys. Rev. E {\bf 91}, 023208(2015).

\bibitem{Daraio2016PRE93.052203}C. Chong, E. Kim, E. G. Charalampidis, H. Kim, F. Li, P. G. Kevrekidis, J. Lydon, C. Daraio, and J. Yang, Phys. Rev. E {\bf 93}, 0052203(2016).

\bibitem{Pozharskiy2015PRE92.063302}D. Pozharskiy, Y. Zhang, M. O. Williams, D. M. McFarland, P. G. Kevrekidis, A. F. Vakakis, and I. G. Kevrekidis, Phys. Rev. E  {\bf 92}, 063203(2015).

\bibitem{Zhang2017Exp57.505}Y. Zhang, D. Pozharskiy, D. M. McFarland, P. G. Kevrekidis, I. G. Kevrekidis, A. F. Vakakis, Exp. Mech. {\bf 57}, 505(2017).
\bibitem{Kovaleva2018PRE98.012205}A. Kovaleva, Phys. Rev. E  {\bf 98}, 012205(2018).
\bibitem{Kadanlof1995PRL74.1268}Y. Du, H. Li, and L. Kadanoff, Phys. Rev. Lett. {\bf 74}, 1268(1995).

\bibitem{Senf1995PRL74.2686}R. S. Sinkovits, and S. Sen, Phys. Rev. Lett. {\bf 74}, 2686(1995).

\bibitem{Luding1994PRE50.4113}S. Luding, E. Cl{\'e}ment, A. Blumen, J. Rajchenbach and J. Dyran, Phys. Rev. E {\bf 50}, 4113(1994).

\bibitem{Lanndau1959}L. D. Landau, and E. M. Lifshitz, {\em Theory of Elasticity}, (Butterworth Heinemann, Oxford, 1986).

\bibitem{PRE2006Huang}D. C. Huang, G. Sun, and K. Q. Lu, Phys. Rev. E {\bf 74}, 061306(2006).

\bibitem{PRE2012Huang}D. C. Huang, M. Lu, G. Sun, F. D. Feng, M. Sun, H. P. Wu, and K. M. Deng, Phys. Rev. E {\bf 85}, 031305(2012).

\bibitem{PT2022Huang}C. H. Li, X. Li, T. F. Jiao, F. L. Hu, M. Sun,  and D. C. Huang, Powder. Tech. {\bf 401}, 117271(2022).

\bibitem{PLA2017Huang}L. Ma, D. C. Huang, W. Z. Chen, T. F. Jiao, M. Sun, F. L. Hu, and J. Y. Su, Phys. Lett. A {\bf 381)}, 542(2017).

\end{thebibliography}
\end{document}